\documentclass[journal]{IEEEtran}

\usepackage{xcolor,soul,framed} 

\colorlet{shadecolor}{yellow}
\usepackage[pdftex]{graphicx}
\graphicspath{{../pdf/}{../jpeg/}}
\DeclareGraphicsExtensions{.pdf,.jpeg,.png}

\usepackage[cmex10]{amsmath}
\usepackage{array}
\usepackage{mdwmath}
\usepackage{mdwtab}
\usepackage{eqparbox}
\usepackage{url}
\usepackage{booktabs} 
\usepackage{float}
\usepackage{placeins}
\usepackage{afterpage}

\newcommand{\thickhline}{\noalign{\hrule height 2pt}}

\begin{document}
    \title{AlzhiNet: Traversing from 2DCNN to 3DCNN, Towards Early Detection and Diagnosis of Alzheimer's Disease}
\author{Romoke~Grace~Akindele$^{1}$,~Samuel~Adebayo$^{2}$,~\IEEEmembership{Member,~IEEE,}~
        Paul~Shekonya~Kanda$^{1}$, and~Ming~Yu$^{1,3}$,~\IEEEmembership{Member,~IEEE}%
\IEEEcompsocitemizethanks{
\IEEEcompsocthanksitem $^{1}$School of Electronics and Information Engineering, Hebei University of Technology, Tianjin, 300401, China.\protect\\
E-mail: 201940000019@stu.hebut.edu.cn, 201641401001@stu.hebut.edu.cn
\IEEEcompsocthanksitem $^{2}$At the time of this research, Samuel Adebayo was a PhD student at the School of Electronics, Electrical Engineering and Computer Science, Queen's University Belfast, Belfast, BT9 5BN, United Kingdom.\protect\\
E-mail: sadebayo01@qub.ac.uk
\IEEEcompsocthanksitem $^{3}$School of Artificial Intelligence, Hebei University of Technology, Tianjin, 300401, China.\protect\\
E-mail: yuming@hebut.edu.cn
}
}



\markboth{IEEE TRANSACTIONS ON BIOMEDICAL ENGINEERING (TBME), VOL.~60, NO.~12, OCTOBER~2024
}{Roberg \MakeLowercase{\textit{et al.}}: High-Efficiency Diode and Transistor Rectifiers}

\maketitle

\begin{abstract}
Alzheimer's disease (AD) is a progressive neurodegenerative disorder with increasing prevalence among the aging population, necessitating early and accurate diagnosis for effective disease management. In this study, we present a novel hybrid deep learning framework that integrates both 2D Convolutional Neural Networks (2D-CNN) and 3D Convolutional Neural Networks (3D-CNN), along with a custom loss function and volumetric data augmentation, to enhance feature extraction and improve classification performance in AD diagnosis. According to extensive experiments, AlzhiNet outperforms standalone 2D and 3D models, highlighting the importance of combining these complementary representations of data. The depth and quality of 3D volumes derived from the augmented 2D slices also significantly influence the model's performance. The results indicate that carefully selecting weighting factors in hybrid predictions is imperative for achieving optimal results. Our framework has been validated on the Magnetic Resonance Imaging (MRI) from Kaggle and MIRIAD datasets, obtaining accuracies of \(98.9\%\) and \(99.99\%\), respectively, with an AUC of \(100\%\). Furthermore, AlzhiNet was studied under a variety of perturbation scenarios on the Alzheimer's Kaggle dataset, including Gaussian noise, brightness, contrast, salt and pepper noise, color jitter, and occlusion. The results obtained show that AlzhiNet is more robust to perturbations than ResNet-18, making it an excellent choice for real-world applications. This approach represents a promising advancement in the early diagnosis and treatment planning for Alzheimer's disease.
\end{abstract}


%
\IEEEpeerreviewmaketitle


\section{Introduction}

\IEEEPARstart{R}{ecognizing} the need for immediate action to alleviate the impact of Alzheimer's disease, a costly disease with few treatment options, there has been a shift towards identifying people in the early stages of the disease. Not all people with mild cognitive impairment will progress to dementia, and while there is no current cure, it is crucial to improve diagnostic rates to identify those at the highest risk early on and implement measures to mitigate further progression \cite{rasmussen2019alzheimer}. The prevalence of AD is expected to rise significantly, with a projected 14 million affected individuals by 2050 which necessitates early diagnosis, which is crucial for effective disease management \cite{fulton2019classification}.

\begin{figure}[h]
    \centering
    \includegraphics[width=0.6\columnwidth]{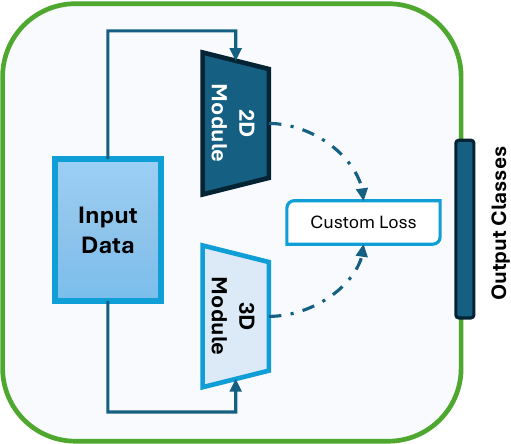}
    \caption{High-level overview of AlzhiNet Architecture. The framework processes input data through 2D and 3D modules, integrating their outputs via a custom loss function to produce the final output classes.}
    \label{fig:highlevel_overview}
\end{figure}

A variety of computer-aided diagnosis (CAD) approaches have been proposed for the early diagnosis of various stages of AD using structural Magnetic resonance imaging (sMRI). Accurate diagnostic methods, including Positron emission tomography (PET) scans, Magnetic resonance imaging (MRI), and Single Photon Emission Computed Tomography (SPECT) scans, are crucial for understanding and treating Alzheimer's disease, with MRI being particularly useful for studying AD-related brain changes \cite{lian2018hierarchical} \cite{zhang2021explainable} \cite{aberathne2023detection}. The utilization of MRI is an efficient way to study brain changes related to Alzheimer's disease while the patient is alive. Unlike X-rays, MRI does not emit ionizing radiation, making it a valuable tool for tracking the progression of Alzheimer's disease and monitoring the effectiveness of treatment \cite{wen2020convolutional}. However, accurate image information is essential for making decisions throughout the patient care process, from detection and characterization to treatment response assessment and disease monitoring. So far, deep learning has played a crucial role in guiding interventional procedures, surgeries, and radiation \cite{chan2020deep}. A lightweight three-layer 3D convolutional network Module (3D-M) was proposed by \cite{wang2023hybrid} for HSIs' spectral-spatial feature extraction. In their work, a hybrid network model (HNM) was presented, extracting spectral-spatial features with low computational complexity and leveraging structural information. Furthermore, \cite{yeoh2023transfer} suggested filling the void by leveraging transfer learning of 2D pre-trained weights in the 3DCNN in medical imaging, explicitly for diagnosis of knee osteoarthritis (OA) with a focus on binary classification for knee recognition with OA. Alzheimer's disease (AD), is a neurodegenerative illness that affects brain cells gradually. It's imperative to catch this early to avoid irreversible memory loss.
This necessitates the need to navigate from 2D Convolutional Neural Network (2DCNN) to 3D Convolutional Neural Network (3DCNN), which involves extending the convolutional neural network architecture from processing two-dimensional data to three-dimensional data. 3DCNN leverages the advantages of using both spatial and temporal feature extraction to capture relevant information from the images. By making use of a 2D convolutional kernel in 2DCNN, it only gathers spatial data in two dimensions; it ignores three-dimensional data. In comparison to the 2DCNN, the 3DCNN can capture spatial information in all three dimensions by incorporating depth (time) into the convolutional process. 3D Convolutional Neural Networks are capable of comprehending both spatial and temporal data. Even between different 2D slices, this enables the 3DCNN to provide a more comprehensive view of volumetric data. As a result, the 3DCNN can extract more distinguishable representations from the data compared to the 2DCNN \cite{klaiber2021systematic}.

The remainder of this work consists of the following sections: Section 2 provides a summary of recent advances in convolutional neural networks (CNNs), including pre-trained CNNs for feature extraction based on ResNet-18, 2D-CNN models, and 3D-CNN models. The research contributions are detailed in Section 3. Section 4 describes the study's methodology. Section 5 is about the experimental and system setup. Section 6 summarizes the study's findings. Lastly, the paper concludes with a discussion of the implications of the findings and recommendations for future research in Section 7.

\section{Related Work}
\subsection{Pre-trained CNNs for feature extraction based on ResNet-18}

Recently, pre-trained CNNs have demonstrated potential in diagnosing cognitive illnesses from brain MR images. Prominent deep neural networks like Alex-Net \cite{shakarami2020cad}, VGG-16 \cite{qiu2020development}, ResNet-34  \cite{amin2019alzheimer}, ResNet-50 \cite{jabason2019classification}, Squeeze-Net, and InceptionV3 \cite{odusami2021comparable} have been previously trained and utilized for MRI data analysis. The authors in \cite{he2016deep} developed ResNet-18, an 18-layer deep CNN model, which was pre-trained on more than a million images from the ImageNet database. Employing a pre-trained network has the added advantage of enhancing the model's performance even with fewer training samples. A novel neural network-based breast cancer diagnostic method was presented by \cite{yu2019abnormality}, such that various variants of ResNet, which includes ResNet-18, ResNet-50, and ResNet-101, were pre-trained to extract features from ROIs. With a mean classification accuracy of 95.91 \%, ResNet-18 emerged as the most promising model in the system's ability to classify mammography pictures into normal and abnormal regions. Odusami et al. \cite{odusami2021analysis} presented a deep learning-based strategy to predict AD, early MCI (EMCI), late MCI (LMCI), and mild cognitive impairment (MCI). With the optimized ResNet18 network, classification accuracy on EMCI vs. AD, LMCI vs. AD, and MCI vs. EMCI situations was 99.99 \%, 99.95 \%, and 99.95 \% on the functional Magnetic Resonance Imaging (fMRI) used. Imaging tools for AD diagnosis have improved with the use of deep learning algorithms. However, due to closely connected aspects in brain structure, multi-class categorization is still a challenge \cite{odusami2021resd}. Therefore, a hybrid ResD method based on Resnet-18 and Densenet-121 was employed for AD multiclass classification using an MRI dataset. Their experiments demonstrated that the hybrid model being proposed performs better than other techniques that have been previously used or studied.

\subsection{2D-CNN Models}

Two-dimensional (2D) CNN has been effectively utilized in different areas, including image classification, face recognition, and natural language processing. To enhance the efficacy of deep learning methods for the diagnosis of Alzheimer's disease, three approaches were suggested to utilize 2D CNNs on 3D MRI data, \cite{liang2021alzheimer}. Validated on the Alzheimer's Disease Neuroimaging Initiative (ADNI) dataset, the results show that the suggested strategy greatly improves model performance in the field of medical imaging. Validated on the Alzheimer's Disease Neuroimaging Initiative (ADNI) dataset, the results show that the suggested strategy greatly improves model performance in the field of medical imaging. Although medical imaging systems for AD have improved with the application of machine learning and deep learning techniques, multi-class classification remains difficult because of highly related features. For an accurate diagnosis, a two-dimensional deep convolutional neural network (2D-DCNN) is proposed using an unbalanced MRI dataset \cite{nawaz2020deep}. The model outperforms the latest state-of-the-art methods with imbalanced classes, achieving 99.89 \% classification accuracy. 2D CNNs can identify temporal dependencies between 2D image slices in a 3D MRI image volume, but they are not able to identify spatial dependencies within an image slice. Hence, by simulating the sequence of MRI features generated by a CNN and deep sequence-based networks for AD detection, the authors in \cite{ebrahimi2021deep} proposed a solution. This study attempts to address this problem.

\subsection{3D-CNN Models}

In \cite{parmar2020deep}, the researchers detail a method of utilizing resting state fMRI data and a 3D CNN to predict the advancement of AD in an individual patient. This technique gathers spatial information from a four-dimensional volume, streamlining the feature extraction process. The study demonstrates that a simple deep learning model was able to classify AD with 94.58 \% accuracy. Three effective techniques have been developed by \cite{yang2018visual} to produce visual explanations for Alzheimer's disease classification using 3D convolutional neural networks (3D-CNNs). The first one segments 3D images hierarchically using sensitivity analysis, and the second one shows network activations on a geographical map. While all methods identify relevant brain regions for diagnosing Alzheimer's disease, the sensitivity analysis method has difficulties with the cerebral cortex's loose distribution. Study \cite{qiao2021fusion} combines multi-view 2D and 3D convolutions to offer a unique classification approach for MRI-based Alzheimer's disease diagnosis. The technique uses MRI to obtain global subject-level data after extracting local slice-level features from each slice. Their combination of worldwide and local data enhances discriminative capabilities when experimented on the ADNI-1 and ADNI-2 datasets. According to \cite{hosseini2016alzheimer}, it suggests utilizing brain scans to detect Alzheimer's disease (AD) using a deep 3D convolutional neural network (3D-CNN) with the capability of adapting to various datasets and learning generic characteristics that capture AD biomarkers. Their 3D-CNN, built on a pre-trained 3D convolutional autoencoder, is tailored for task-specific classification, as demonstrated in experiments on the CAD Dementia MRI dataset.

Though a series of studies have been carried out using 2D-CNN and 3D-CNN independently, they are not enough to provide a comprehensive analysis of spatial information and enhance contextual understanding, thereby causing information loss. So, it is imperative to capture spatial relationships and context in medical image data for accurate analysis and diagnosis, even though the temporal dimension may not apply to static medical images. The importance of early detection cannot be overstated in initiating interventions and treatments that can slow down the disease's progression. Moreover, healthcare providers can tailor treatment plans to an individual's specific disease progression if they understand the spatio-temporal patterns of brain changes associated with Alzheimer's disease. As a result, our approach aims to provide a deeper understanding of Alzheimer's disease by shifting focus to a spatio-temporal perspective. The study provides insight into how Alzheimer's disease progresses over time as well as how certain brain regions are affected at various stages with the integration of the two models. This valuable information can be used to monitor disease progression, assess treatment effectiveness, and predict future outcomes with MRI.

Some current methods struggle with noise, impacting classification accuracy and prediction correctness. To address this, we developed 3D models that incorporate noise as part of the augmentation process to effectively learn and classify it. This augmentation occurs during training but is not used during testing. Even in the presence of noise, our model's loss function ensures accurate classification and consistent predictions.

\section{Contributions}
In this study, we propose a novel hybrid deep learning framework that combines a 2D CNN based on ResNet-18 with a 3D-CNN for the classification of Alzheimer's disease. The key contributions of our research are as follows:

\begin{enumerate}
    \item We developed a model architecture that integrates a 2D-CNN and a 3D-CNN architecture, leveraging the strengths of both architectures to process 2D medical imaging data effectively. This combination allows for enhanced feature extraction and robust classification performance.

    \item We introduced a technique to form 3D volumes from 2D images by applying a series of image augmentation functions. This approach ensures that the 3D CNN receives volumetric input data, which is crucial for capturing spatial and volumetric information.

    \item We introduced a custom combined loss function that strategically combines the cross-entropy loss for 2D and 3D outputs with a mean squared error (MSE) loss for the softmax probabilities.

    \item Our model demonstrates improved performance in Alzheimer's disease classification by effectively utilizing both 2D and 3D data inputs. The integration of the 3D CNN to weight the 2D CNN enhances the model's ability to capture spatial and volumetric information critical for accurate diagnosis.

    \item We conducted extensive experiments to evaluate the performance of our hybrid model. Our results show that the proposed model outperforms traditional 2D and 3D CNN approaches in terms of accuracy, precision, recall, and F1 score.
    
    \item We performed an ablation study to analyze the impact of different components of our framework; this provides insights into the contribution of each part to the overall model's performance.

    \item The proposed framework is validated on real-world medical imaging data, demonstrating its practical applicability and potential for clinical use. Our approach shows promise for improving early diagnosis and treatment planning for Alzheimer's disease.
\end{enumerate}

By combining the strengths of 2D and 3D CNNs and introducing a novel loss function, our research contributes to advancing the state-of-the-art in Alzheimer's disease classification. It paves the way for future studies to build upon our findings.
\section{Methodology} \label{sec:Methodology section}
In this section, we will detail the methodology used in our study, including data preprocessing techniques, the architecture of the AlzhiNet model, and the custom loss function employed to optimize performance.  Our approach integrates advanced image processing techniques and a hybrid neural network architecture to achieve robust and accurate classification of Alzheimer's disease from 2D medical imaging data. 

\subsection{Model Architecture}
\begin{figure*}[!ht]
    \centering
    \includegraphics[width=1.\textwidth]{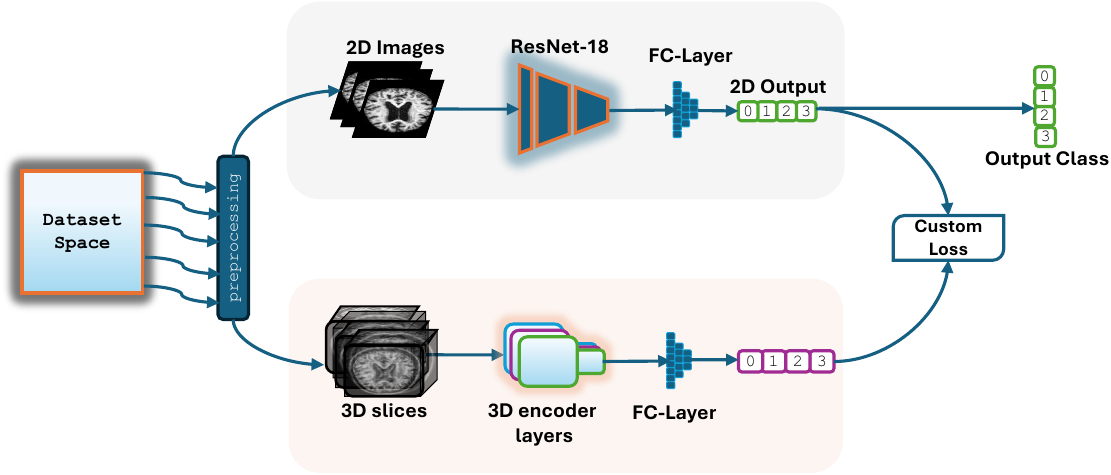}
    \caption{Detailed architecture of AlzhiNet hybrid model. The framework processes both 2D images through ResNet-18 and 3D volumes through 3D encoder layers, integrating their outputs via a custom loss function to produce the final classification output.}
    \label{fig:alzhiNet-architecture}
\end{figure*}

Our novel framework consists of two major modules: the 2D module and the 3D module. These modules work together to leverage the strengths of both 2D and 3D CNN for the classification of Alzheimer's disease. The detailed architecture is illustrated in Figure \ref{fig:alzhiNet-architecture}. We detail the architecture of both module in Sections \ref{subsubsec: 2D-Module} and \ref{subsubsec: 3D-Module}

\subsubsection{2D Module} \label{subsubsec: 2D-Module}
The 2D module is based on the ResNet-18 architecture, which is widely recognized for its effectiveness in classification tasks for example, classifying Alzheimer's disease \cite{odusami2021analysis, odusami2021resd, yu2019abnormality}. This module utilizes a pre-trained ResNet-18 model with weights initialized from the pre-trained ResNet-18 on the ImageNet dataset; however, the final fully connected layer of the ResNet-18 is replaced with a new fully connected layer that outputs the desired number of classes for Alzheimer's disease classification. The architecture and flow of the 2D module include:

\begin{itemize}
    \item Input Layer: Takes in 2D slices of medical images.
    \item Convolutional Layers: Multiple layers with residual connections to extract high-level features.
    \item Fully Connected Layer: Adjusted to output the number of classes for classification.
    \item Activation Function: ReLU (Rectified Linear Unit) is used throughout the network to introduce non-linearity.
\end{itemize}

\subsubsection{3D Module} \label{subsubsec: 3D-Module}
In parallel to the 2D module, the 3D encoder module is designed to handle volumetric data by processing stacked 2D augmented images to form a 3D volume. This module consists of a series of 3D convolutional layers followed by batch normalization and pooling layers to capture spatial and volumetric features critical for accurate classification. As depicted in Table~\ref{Table1:Architecture of 3D Module}, the architecture of the 3D module includes:

\begin{itemize}
    \item Input Layer: Takes in 3D volumes created from augmented 2D images.
    \item 3D Convolutional Layers: Three 3D convolutional layers with kernel size \(3 \times 3 \times 3\), stride 1, and padding 1 to extract volumetric features.
    \item Batch Normalization: Applied after each convolutional layer to stabilize and accelerate training, refer to Table \ref{Table1:Architecture of 3D Module} for details on the 3D Module Encoder.
    \item Activation Function: ReLU is used after each convolutional layer.
    \item Pooling Layers: Average pooling layers to reduce dimensionality while preserving important features.
    \item Fully Connected Layer: A fully connected layer followed by a final output layer that predicts the class probabilities.
\end{itemize}

\begin{table}[h!]
\centering
\scriptsize 
\caption{Architecture of 3D Encoder Module.}
\label{Table1:Architecture of 3D Module}
\begin{tabular}{@{}lccc@{}}
\toprule
\textbf{Type} & \textbf{In-Channels} & \textbf{Out-Channels} & \textbf{K-Size} \\
\midrule
3D Conv 1 & 3 & 64 & 3 \\

Batch Norm & - & 64 & - \\

3D Conv 2 & 64 & 128 & 3 \\

Batch Norm & - & 128 & - \\

3D Conv 3 & 128 & 256 & 3 \\

Batch Norm & - & 256 & - \\

Average Pool & - & - & 3 \\

Adaptive Average Pooling 3D & - & - & 1 \\

Dense & 256 & 512 & - \\

Dense & 512 & num of classes & - \\
\bottomrule
\end{tabular}
\end{table}
\subsection{Hybrid Model Integration and Custom Loss Function}
The integration of the 2D and 3D modules in AlzhiNet involves combining their respective outputs to make the final prediction. The 2D module processes 2D images, while the 3D module processes corresponding 3D volumes. The final prediction is achieved by combining the outputs from both modules and using a custom loss function (as depicted in Section \ref{subsec: Custom Loss Function} that ensures the 3D module refines the 2D model's predictions.

Let \( O_{2d} \) and \( O_{3d} \) be the output logits from the 2D and 3D modules, respectively, and let \( S_{2d} \) and \( S_{3d} \) be their corresponding softmax probabilities. The overall hybrid model prediction \( O_{h} \) is given by:

\begin{equation}
    O_{h} = \alpha \times O_{2d} + \beta \times O_{3d}
    \label{eq:Ohybrid}
\end{equation}

where \( \alpha \) and \( \beta \) are weighting factors that balance the contributions of the 2D and 3D modules, respectively.

\subsubsection{Custom Loss Function} \label{subsec: Custom Loss Function}
The custom loss function is a critical component of our hybrid model, ensuring accurate classification and consistency of model predictions. This function is designed to leverage the complementary strengths of the 2D and 3D modules, it combines the cross-entropy loss for 2D and 3D outputs with a mean squared error (MSE) loss for the softmax probabilities. The formulation of the custom loss function is broken down as follows:

Firstly, the cross-entropy loss for the 2D model outputs \(O_{2d}\) and the target classes \(T\) is given by:

\begin{equation}
    \label{eq:L2d}
    L_{2d} = L_{ce2d}(O_{2d}, T)
\end{equation}

Next, the cross-entropy loss for the 3D model outputs \(O_{3d}\) and the target classes \(T\) is defined as:

\begin{equation}
    \label{eq:L3d}
    L_{3d} = L_{ce3d}(O_{3d}, T)
\end{equation}

Finally, the complete custom loss function incorporates the above components along with the mean squared error loss between the softmax probabilities \(S_{2d}\) of the 2D model and \(S_{3d}\) of the 3D model:

\begin{equation}
    \label{eq:Lcustom}
    L_{cm} = L_{2d} + L_{3d} + \lambda \times L_{mse}(S_{2d}, S_{3d})
\end{equation}

In this formulation, \(L_{2d}\) (Equation \ref{eq:L2d}) and \(L_{3d}\) (Equation \ref{eq:L3d}) represent the cross-entropy losses \footnote{For the dataset with binary class, we used Binary Cross Entropy.} for the 2D and 3D models, respectively. $\lambda$ is a balancing factor that controls the influence of the MSE loss on the total loss. Thus, the term \(\lambda \times L_{mse}(S_{2d}, S_{3d})\) in Equation \ref{eq:Lcustom} balances the mean squared error loss with the cross-entropy losses to ensure consistency between the softmax probabilities of the 2D and 3D models.

This custom loss function ensures that the model focuses on accurate classification through the cross-entropy loss and also maintains consistency between the 2D and 3D model predictions through the MSE loss. By combining these loss components, our model achieves robust performance and improves the classification accuracy for Alzheimer's disease.

\section{Experimental and system Setup}
\begin{figure*}
\scriptsize
\centering
\begin{tabular}{c c c c}
\toprule%
$Mild Demented$ &  $Moderate Demented$ & Non Demented & Very Mild Demented  \\
\midrule
{\includegraphics[width=0.2\textwidth]{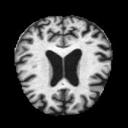}}   & {\includegraphics[width=0.22\textwidth]{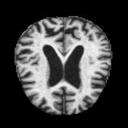}} & {\includegraphics[width=0.2\textwidth]{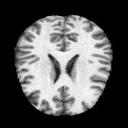}} & {\includegraphics[width=0.22\textwidth]{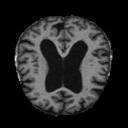}} \\

$\textbf{$ $Class-0}$ & $\textbf{$ $Class-1}$ & $\textbf{$ $Class-2 }$& $\textbf{$ $Class-3}$ \\
\bottomrule
\end{tabular}
\caption{Visual depiction of standard brain MRI samples from Kaggle.\label{figdatasetvid}}
\end{figure*}

This section outlines the experimental setup used to evaluate the performance of our hybrid model. We describe the datasets, preprocessing steps, training procedures, and evaluation metrics used in our experiments.
The typical Kaggle brain MRI samples for each class are shown in Fig.~\ref{figdatasetvid}.

\subsection{Setup of the Experiment}
The experiment was conducted on an Alienware m16 Linux desktop configured with a core i9 13900HX processor, 32.0GB RAM, and Nvidia GeForce RTX 4070 16GB GPU for training. Also, all images in the experiment were 224 x 224 pixels. other hyperparameters used in this study have been stated in Table~\ref{Table1:Architecture of 3D Module}.

\begin{figure}[!b]  
    \scriptsize  
    \centering  
    \begin{tabular}{@{}c@{}c@{}c@{}}  
        \includegraphics[width=0.12\textwidth]{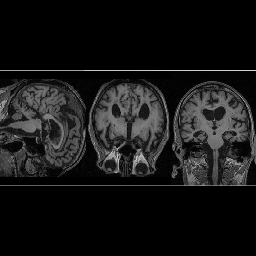} &   
        \includegraphics[width=0.12\textwidth]{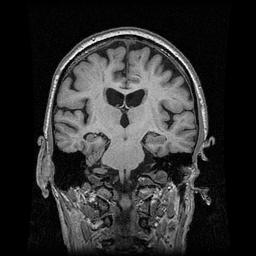} &   
        \includegraphics[width=0.12\textwidth]{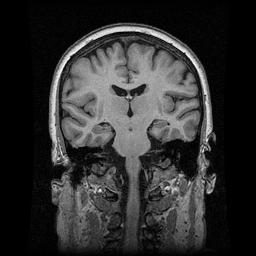} \\
        
        \textbf{a. Sagittal, Axial,}\\ \textbf{Coronal} &   
        \textbf{b. CN} &   
        \textbf{c. AD}  
    \end{tabular}  
    \caption{Visualization of Brain MRI Samples: (a) Different Views of MIRIAD MRI Images (b) Normal Control-CN (c) Alzheimer’s Disease (AD) brain imaging.}  
    \label{fig:brain_mri_samples}
\end{figure}

\subsection{Datasets}
\subsubsection{Kaggle Alzheimer's Dataset}
This paper utilizes Alzheimer-MRI data sets obtained from multiple websites, hospitals, and public repositories via KAGGLE. The data consists of preprocessed whole-brain T1-Weighted MRI Images that have been originally resized to 128 x 128 pixels. The KAGGLE data sets include 6400 axial images from individuals aged 18 to 96, encompassing the full spectrum of Alzheimer's disease development \cite{dubey2019alzheimer}. Please refer to Table~\ref{tableclasses1} for the dataset breakdown.

\begin{table}[h]
\centering
\caption{Number of images in the AD Dataset on Kaggle.} \label{tableclasses1}
\begin{tabular}{@{}lll@{}}
\toprule
 Dementia Level & Classes & No. of images \\
  &  &  present  \\
\midrule
Mild	&  0  & 896	\\
Moderate	&  1  & 64	\\
Non	&  2  & 3200	\\
Very Mild	&  3  & 2240	\\
 \bottomrule 
\end{tabular}
\end{table}

Deep learning models often face class imbalance in medical imaging datasets due to the low occurrence of specific disorders. This results in a different number of samples in each class, which can negatively impact machine learning models' performance and lead to biased predictions. To address this, standard methods like Data Augmentation, Ensemble Methods, Cost-sensitive Learning, Synthetic Data Generation, Data Resampling, Weighted Loss Function, and adjusting evaluation measures like accuracy, recall, and F1-score are used. To enhance minority class samples, to be precise moderately demented, an augmentation was performed and the detail is given in section 6.2. A 70:30 split was applied to the dataset, and by creating slightly altered duplicates of the current minority class samples, the size of the minority class was increased. The overall size of the dataset on the desktop is 26.8 MB.

 Table~\ref{tabledataset} highlights the increased number of samples achieved through augmentation.

\begin{table}[h]
\centering
\caption{Dataset distribution used in this study.}\label{tabledataset}
\begin{tabular}{@{}lllll@{}}
\toprule
 & Mild  & Moderate  & Non & Very Mild\\
\midrule
Train & 627	& 480 & 2240 & 1568\\
Test & 269 &  27 &	960	& 672\\
 \bottomrule 
\end{tabular}
\end{table}

\subsubsection{MIRIAD Dataset}

Also included in the experiments was a public dataset of Minimal Interval Resonance Imaging in Alzheimer's Disease (MIRIAD) \cite{malone2013miriad}. This collection contains information on 46 individuals with Alzheimer's disease (AD) and 23 individuals with control normal (CN), shown in Fig.~\ref{fig:brain_mri_samples} along with their three planes. MRI brain scans are compiled from Alzheimer's patients and healthy elderly people. The purpose of collecting a large number of MRI scans from each participant over 2 weeks to 2 years was to determine whether MRI could serve as an outcome measure for Alzheimer's clinical trials. 708 MRI scans were taken in total from CN and AD subjects as represented in Table~\ref{tableclasses}. A 1.5 T Sigma MRI scanner was used for all MRI scans performed by the same radiographer. Three-dimensional T1-weighted images were collected using an IR-FSPGR (inversion recovery prepared fast spoiled gradient recalled) sequence. This experiment consisted of a 24-cm field of view, a 256-x-256 matrix, 124 1.5 mm coronal partitions, a 15 ms TR, 5.4 ms TE, a 15° flip angle, and a 650 ms TI. First, we converted the NII data into MRI images using the med2image approach \cite{med2image}, viewed on July 23, 2024, and the selected axial planes. In Fig.~\ref{fig:brain_mri_samples}, we can see the visual contrast between CN and AD, and Table~\ref{tabledataSPLIT} shows the number of samples acquired after the MIRIAD Dataset was split.

\begin{table}[h]
    \centering
    \caption{Samples in the MIRIAD.} \label{tableclasses}  
    \begin{tabular}{@{}ll@{}}  
    \toprule  
    \textbf{CDR} & \textbf{MIRIAD Images Present} \\
    \midrule  
    0 Control Normal & 243 \\
    1 Alzheimer’s Disease & 465 \\
    TOTAL & 708 \\
    \bottomrule  
    \end{tabular}  
\end{table}  

\begin{table}[h]
\centering
\caption{Number of Samples obtained after splitting.}\label{tabledataSPLIT}
\begin{tabular}{@{}llll@{}}
\toprule
 CDR & Value & MIRIAD & Train-test split \\ 
\midrule
Control & 0 & 243  &Train = 170 \\
Normal &  &  & Test = 73 \\ \addlinespace[1pt] 
Alzheimer's Disease & 1	& 465  & Train = 325\\
 & 	&   & Test = 140\\ \addlinespace[1pt]
\bottomrule
\end{tabular}
\end{table}

\subsection{Dataset Preprocessing}
Preprocessing is essential to improving the image data by eliminating distortions and enhancing crucial features for further processing. To improve model performance and generalization, we applied image data augmentation, incorporating various augmentation functions to strengthen the diversity of the training data.

\textbf{Data Augmentation and 3D Volume Formation:}
Considering that 3D CNNs require volumetric (3D) data to fully leverage their spatial hierarchical feature extraction capabilities and that the Kaggle and MIRIAD datasets primarily comprise 2D images, we developed a novel technique to transform these 2D images into 3D volumes. This method involves applying a set of data augmentations to each 2D image, thereby generating multiple augmented versions that are then stacked to form a 3D volume.

\begin{figure*}
    \centering
    \includegraphics[width=0.9\textwidth]{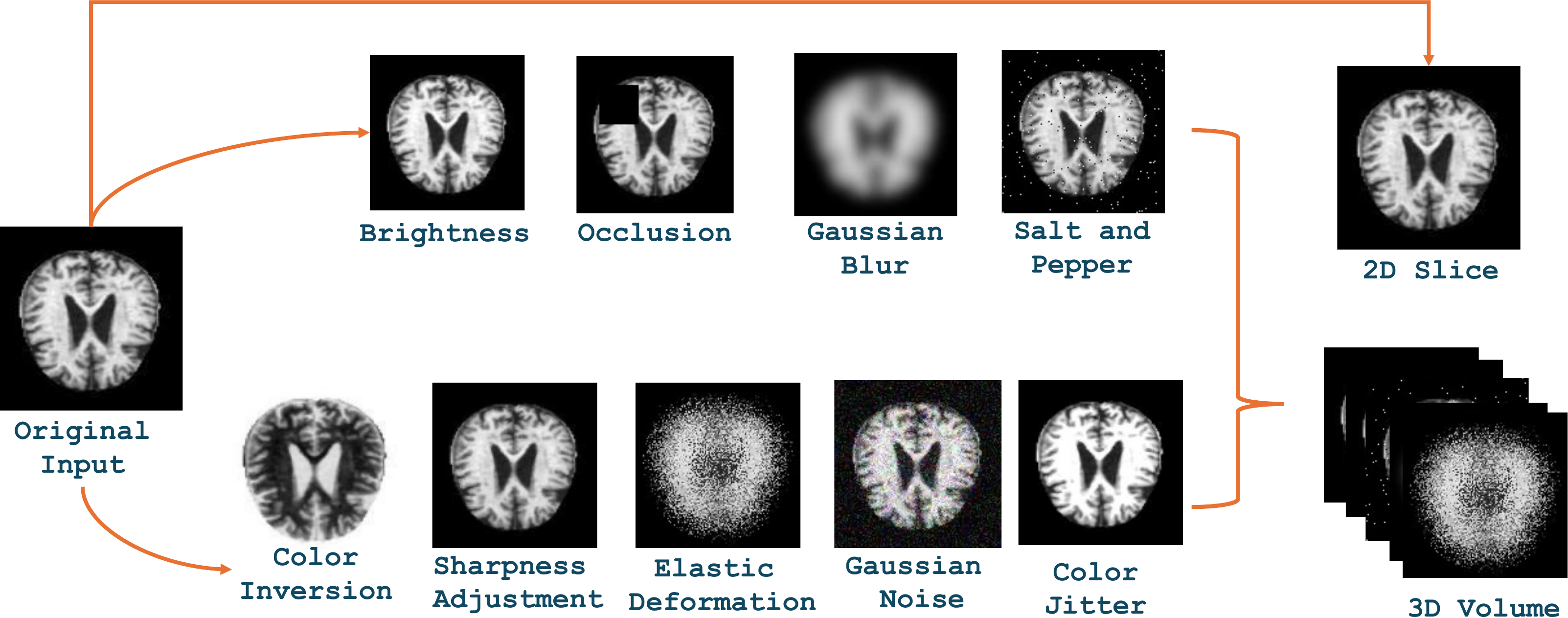}
    \caption{Generation of Augmented 3D volumes from 2D Brain Imaging Slices using diverse data Augmentation Techniques}
    \label{fig:alzhiNet-dataset}
\end{figure*}

Figure \ref{fig:alzhiNet-dataset} illustrates the transformation of a single 2D brain imaging slice through multiple data augmentation techniques, resulting in a set of augmented images. These images are subsequently stacked to construct a 3D volume. Each panel depicts a distinct augmentation applied to the original image, showcasing the preprocessing steps employed to enrich the AlzhiNet architecture and enhance the robustness and generalization of the model.

The following augmentation techniques were applied to achieve this transformation: elastic deformation, color inversion, sharpness adjustment, salt-and-pepper noise addition, brightness adjustment, color jitter, Gaussian noise addition, Gaussian blur, and occlusion. Each augmented image serves to simulate various realistic conditions that might affect MRI scans, such as motion artifacts or variable scan qualities. This process is mathematically formulated as follows:

\begin{equation} \label{eqn:3D-formation}
    3\mathcal{D}_{vol} = \left[ \text{Aug}_{1}(x), \text{Aug}_{2}(x), \ldots, \text{Aug}_{9}(x) \right]
\end{equation}

where \( \text{Aug}_{i}(x) \) represents the \(i\)-th augmentation function applied to the original 2D image \( x \).

\subsection{Training Procedure}
Given a sample image from Dataset space \(D\), we preprocess it to obtain 2D images and 3D volume slices, as described in the preprocessing section and illustrated in Figure \ref{fig:alzhiNet-architecture}.

The 2D module (ResNet-18) was initialized with pre-trained weights from the ImageNet dataset, while the 3D module was initialized with random weights. The fully connected layers in both modules were adjusted to output the desired number of classes for Alzheimer's disease classification. We used the Adam optimizer with an initial learning rate of \(1 \times 10^{-4}\). The custom loss function described in Section \ref{subsec: Custom Loss Function} was used to train the model. This function combines cross-entropy loss for both 2D and 3D outputs with a mean squared error (MSE) loss for the softmax probabilities. For the Kaggle dataset, a batch size of 8 was used, while for MIRIAD, 4 was used for training, and the model was trained for 30 epochs, with early stopping implemented to prevent overfitting.

During each forward pass, the 2D and 3D modules processed their respective inputs (2D slices and 3D volumes). The custom loss function calculates the combined loss based on the outputs of both modules. Gradients were computed and propagated back through the network during the backward pass, and the optimizer updated the model weights accordingly. After each epoch, the model was evaluated on the validation set (separate from the test set) to monitor performance and adjust learning rates if necessary. After training, only the weights of the ResNet-18 (2D module) were saved, resulting in an improved version of the ResNet-18 model enhanced by the combined learning process with the 3D CNN.

\section{Results}
Our approach was validated based on the following metrics: Confusion matrix, Accuracy, Precision, Recall/Sensitivity, F1-Score, and Specificity. 

Table \ref{tab:alzheimers_metrics} presents the comparisons of our proposed model to the state-of-the-art model displaying the measures for accuracy, precision, recall/sensitivity, F1-score, and specificity that are used to analyze the performance in a generic way. On the Kaggle dataset,the accuracy, precision, recall, F1-score, and specificity values achieved by our technique are 98.76 \%, 99.00 \%, 99.00 \%, 99.00 \%, and 99.20 \% respectively. As compared to our AlzhiNet model,\cite{prasanna2024ad}, \cite{ahmed2022dad}, \cite{abunadi2022deep} achieved slightly better results in their classification tasks. This means that their model was able to identify most of the positive instances accurately in contrast to ours. Notwithstanding, MIRIAD datasets perform exceptionally on all the evaluated metrics compared to the SOTA approaches. On the whole, we have demonstrated the superiority of our proposed framework in terms of results from binary class and multi-class tasks, which is important for identifying Alzheimer's disease stages. It shows that our method could be used to identify AD phases more effectively and assist in tailoring treatment plans for patients.
We present the results in Table \ref{tab:alzheimers_metrics}:

\begin{table*}[htbp]
\centering
\caption{Performance Metrics for Alzheimer's Disease Detection Models}
\scriptsize
\resizebox{\textwidth}{!}{
\begin{tabular}{lcccccccc}
\toprule
\textbf{Model} & \textbf{Dataset} & \textbf{Acc(\%)} & \textbf{Prec(\%)} & \textbf{Rec(\%)}  & \textbf{F1(\%)} & \textbf{Spty (\%)} & \textbf{AUC (\%)} & \textbf{Params (M)} \\
\midrule
DCNN with VGG16 \cite{sharma2022deep} & Kaggle & 90.40 & 90.50 & 90.40 & 90.40 & - & 96.90 & -  \\
Deep-Ensemble \cite{loddo2022deep} & Kaggle & 97.71 & - & 96.67 & - & 98.22 & - & - \\
Lightweight Deep Learning Model \cite{el2023accurate} & Kaggle & 95.93 & 95.93 & 95.98 & 95.90 & - & - & 6.6 \\
Transfer Learning Model \cite{prasanna2024ad} & Kaggle & 98.99 & 99.11 & 99.31 & 98.71 & - & - & - \\
ADASYN Optimized NN \cite{ahmed2022dad} & Kaggle & 99.22 & 99.30 & 99.14 & 99.19 & - & 99.91 & 1.1 \\
Hybrid Learning MRI \cite{abunadi2022deep} & Kaggle & 99.80 & - & 99.75 & - & 100 & 99.94 & - \\
Curvelet Transform-based MRI \cite{chabib2023deepcurvmri} & Kaggle & 98.62 & - & 99.05 & 99.21 & 98.50 & - & 51.797 \\
\textbf{AlzhiNet (ours)} & Kaggle & \textbf{98.76} & \textbf{99.00} & \textbf{99.00} & \textbf{99.00} & \textbf{99.2} & \textbf{100} & \textbf{12.4} \\
\midrule
CNN \cite{de2023prediction} & MIRIAD & 89.00 & - & - & - & 89.00 & 92.0 & - \\
RBMSDL-RF \cite{hassan2024residual} & MIRIAD & 94.27 & 94.50 & 94.50 & 94.50 & - & - & 155.506 \\
RBMSDL-SoftMax \cite{hassan2024residual} & MIRIAD & 99.10 & 99.10 & 99.10 & 99.80 & - & - & 155.506 \\
ResNet50-SoftMax \cite{alsaeed2022brain} & MIRIAD & 96.00 & - & 96.00 & 97.00 & 95.00 & - & - \\
CNN-SVM \cite{silva2019model} & MIRIAD & 96.07 & - & - & - & - & - & - \\
\textbf{AlzhiNet (ours)} & MIRIAD & \textbf{99.99} & \textbf{100} & \textbf{100} & \textbf{100} & \textbf{100} & \textbf{100} & \textbf{12.4} \\
\bottomrule
\end{tabular}
}
\label{tab:alzheimers_metrics}
\end{table*}

\begin{figure}[htb]
    \centering
    \includegraphics[width=.7\columnwidth]{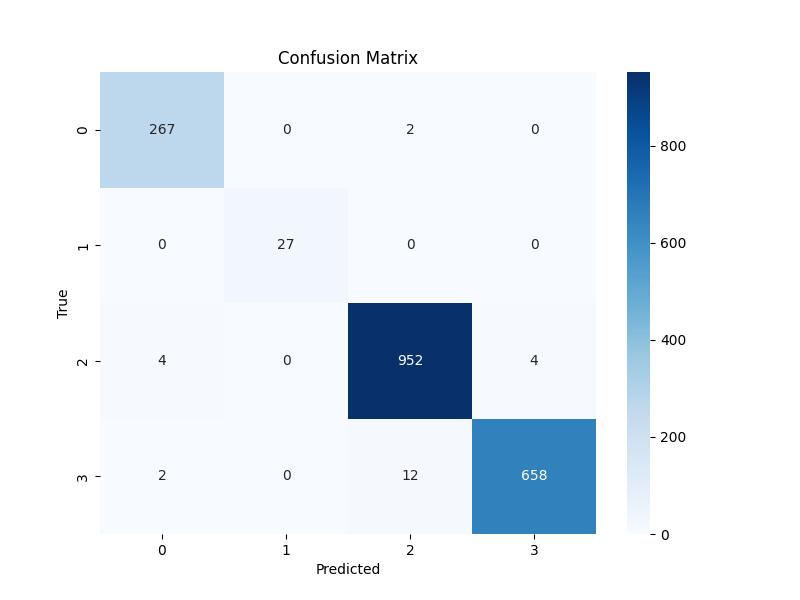}
    \caption{Confusion Matrix plot of AlzhiNet on Kaggle Test Dataset.}
    \label{fig:conf-matrix1}
\end{figure}

\begin{figure}[htb]
    \centering
    \includegraphics[width=.7\columnwidth]{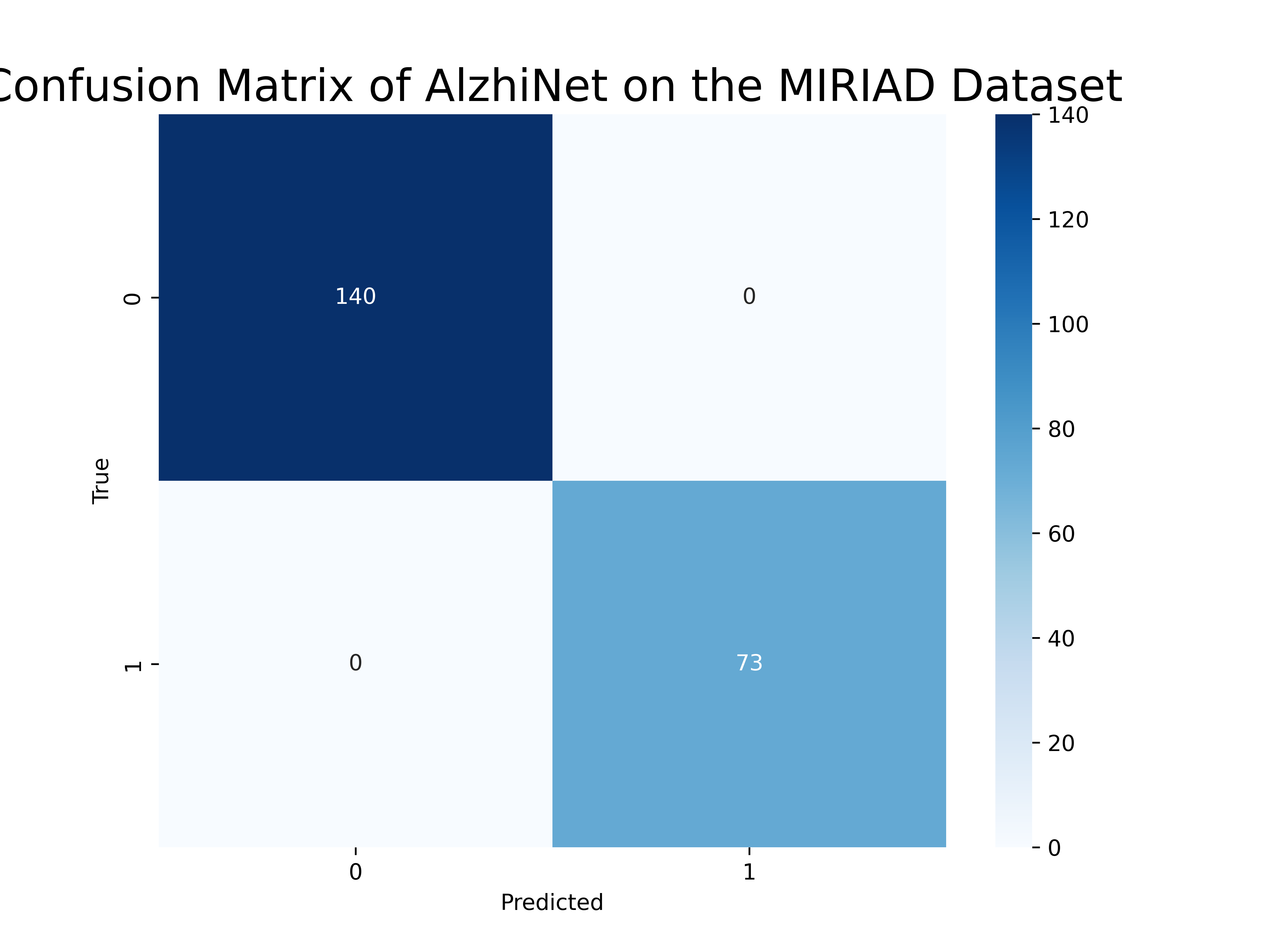}
    \caption{Confusion Matrix plot of AlzhiNet on MIRIAD Test Dataset.}
    \label{fig:conf-matrix}
\end{figure}

Fig.~\ref{fig:conf-matrix1} depicts the visual representation of the AlzhiNet's confusion matrix on the Kaggle Alzheimer's test dataset. Only 8 images in the Non-demented category were misclassified (four as mild dementia and four as very mild dementia). The Very Mild Demented class, out of 672 images, had 14 misclassified (2 as Mild Demented and 12 as Non-Demented). Mild Demented had 2 images misclassified as non-demented. It is noteworthy that all images from the moderate-demented class were accurately classified. The model performed well across all four classes, suggesting that it can predict each class reasonably well. Conversely, Fig.~\ref{fig:conf-matrix} presents the confusion matrix on the MIRIAD dataset, which shows that the two classes were reasonably well predicted by the model, suggesting that our model can cover a wide range of classifications.

\subsection{Ablation Studies}
\begin{table}[!ht]
\centering
\caption{Results of Ablation Studies on the AlzhiNet Model}
\scalebox{0.7}{
\begin{tabular}{lcccccc}
\toprule
\textbf{Experiment} & \textbf{Dataset} & \textbf{Acc (\%)} & \textbf{Prec (\%)} & \textbf{Rec (\%)} & \textbf{F1 (\%)} & \textbf{Param Size (M)} \\
\midrule
2D-Module only & Kaggle & 95.12 & 95.24 & 95.10 & 95.17 & 11.18 \\ 
 & MIRIAD & 98.42 & 98.51 & 98.47 & 98.49 & 11.18 \\
 \midrule
3D-Module Only & Kaggle & 56.75 & 55 & 53 & 54 & 1.22 \\
 & MIRIAD & 61.22 & 65.35 & 58.33 & 57 & 1.22 \\
 \midrule
Without-CustomLoss & Kaggle & 96.88 & 97.01 & 96.89 & 96.95 & 12.4 \\
 & MIRIAD & 99.12 & 99.18 & 99.16 & 99.17 & 12.4 \\
 \midrule
Without-MSE & Kaggle & 97.32 & 97.45 & 97.33 & 97.30 & 12.4 \\
 & MIRIAD & 99.21 & 99.27 & 99.25 & 99.26 & 12.4 \\
 \midrule
$\lambda = 0.1$ & Kaggle & 97.48 & 97.61 & 97.50 & 97.55 & 12.4 \\
 & MIRIAD & 99.45 & 99.51 & 99.49 & 99.50 & 12.4 \\
 \midrule
$\lambda = 0.5$ & Kaggle & 98.76 & 99.00 & 99.00 & 99.00 & 12.4 \\
 & MIRIAD & 99.98 & 100.0 & 100.0 & 100.0 & 12.4 \\
 \midrule
$\lambda = 1.0$ & Kaggle & 98.15 & 98.29 & 98.17 & 98.23 & 12.4 \\
 & MIRIAD & 99.75 & 99.80 & 99.78 & 99.79 & 12.4 \\
 \midrule
9 Augs 3D & Kaggle & 98.76 & 99.00 & 99.00 & 99.00 & 12.4 \\
 & MIRIAD & 99.98 & 100.0 & 100.0 & 100.0 & 12.4 \\
 \midrule
6 Augs 3D & Kaggle & 98.12 & 98.25 & 98.13 & 98.19 & 12.4 \\
 & MIRIAD & 99.74 & 99.80 & 99.78 & 99.79 & 12.4 \\
 \midrule
3 Augs 3D & Kaggle & 97.56 & 97.69 & 97.57 & 97.63 & 12.4 \\
 & MIRIAD & 99.55 & 99.61 & 99.59 & 99.60 & 12.4 \\
\bottomrule
\end{tabular}
}
\label{tab:ablation_results}
\end{table}

To assess the contribution of each component of our proposed AlzhiNet architecture, we conducted extensive ablation studies. These studies were designed to isolate the effects of the 2D and 3D modules, the custom loss function, and various design choices on the overall model performance. We evaluated these components using the Kaggle and MIRIAD datasets, focusing on classification accuracy, precision, recall, F1-score, and model size. The results are summarised in Table \ref{tab:ablation_results}.

\subsubsection{Impact of Individual Modules}

To understand the impact of each module, we first evaluated the 2D and 3D modules independently:

\textbf{2D Module Only (ResNet-18):} We trained the ResNet-18 architecture using 2D slices of the medical images. This model achieved an accuracy of 95.12\% on the Kaggle dataset and 98.42\% on the MIRIAD dataset. While the 2D module alone performs well, it lacks the spatial context provided by the volumetric data.

\textbf{3D Module Only:} The 3D module was trained independently using 3D volumes constructed from augmented 2D slices. This module achieved significantly lower performance, with an accuracy of 56.75\% on the Kaggle dataset and 61.22\% on the MIRIAD dataset. These results suggest that while the 3D module captures essential volumetric features, it struggles to achieve high classification performance on its own. The markedly lower accuracy compared to the 2D module indicates that the spatial context alone is insufficient and must be complemented by the detailed features captured by 2D slices.

\subsubsection{Effect of the Custom Loss Function}
We then evaluated the impact of the custom loss function by conducting experiments with different configurations:

\textbf{Without Custom Loss:} To examine the role of our custom loss function, we trained the model without it, reverting to standard loss components. The results demonstrated a slight drop in performance, with accuracy scores of 96.88\% on Kaggle and 99.12\% on MIRIAD. This indicates that the custom loss contributes positively to the model’s performance, enhancing the consistency between the 2D and 3D module predictions.

\textbf{Without MSE Loss Component:} In this experiment, we removed the MSE component from the custom loss function, using only the cross-entropy losses from the 2D and 3D modules. The performance dropped to 97.32\% on Kaggle and 99.21\% on MIRIAD, indicating that the consistency between the 2D and 3D predictions enforced by the MSE loss is crucial for optimal performance.

\textbf{Varying $\lambda$ Values:} We experimented with different values of the balancing factor $\lambda$ in the custom loss function. We found that a $\lambda$ value of 0.5 provided the best balance, achieving 98.76\% accuracy on Kaggle and 99.98\% accuracy on MIRIAD. Higher or lower values resulted in decreased performance, emphasizing the importance of carefully balancing the contributions of the cross-entropy and MSE losses.

\subsubsection{Impact of 3D Volume Construction}

To evaluate the impact of the 3D volume construction, we varied the number of augmentations used to create the 3D volumes:

\textbf{Different Number of Augmentations:} The construction of 3D volumes was tested with 3, 6, and 9 augmentations to determine the optimal depth for capturing volumetric features. The best performance was observed with 9 augmentations, which achieved an accuracy of 98.76\% on Kaggle and 99.98\% on MIRIAD. Reducing the number of augmentations to 6 resulted in a slight drop in performance, with accuracy falling to 98.12\% on Kaggle and 99.74\% on MIRIAD. The 3 augmentations scenario yielded the lowest performance among the three, indicating that more comprehensive volumetric features are captured when a higher number of augmentations is used. This suggests that the depth and richness of the 3D volumes are important for maximizing the model's ability to generalize and accurately classify medical images.

In summary, the ablation studies on different components of AlzhiNet confirm that the integration of both the 2D and 3D modules, supported by a carefully balanced custom loss function, is key to the success of the AlzhiNet architecture. The 2D module provides detailed feature extraction from the slices, while the 3D module adds crucial spatial context. The custom loss function, particularly the inclusion of the MSE component, ensures coherence between these two modules, resulting in superior performance. The number of augmentations used in constructing 3D volumes also plays an important role, with more augmentations leading to better performance, as evidenced by the results from both the Kaggle and MIRIAD datasets. These findings underline the robustness and efficacy of the AlzhiNet model in handling complex medical imaging tasks.

\subsection{Testing with Noise}
\begin{figure*}[!ht]
    \centering
    \includegraphics[width=0.5\textwidth]{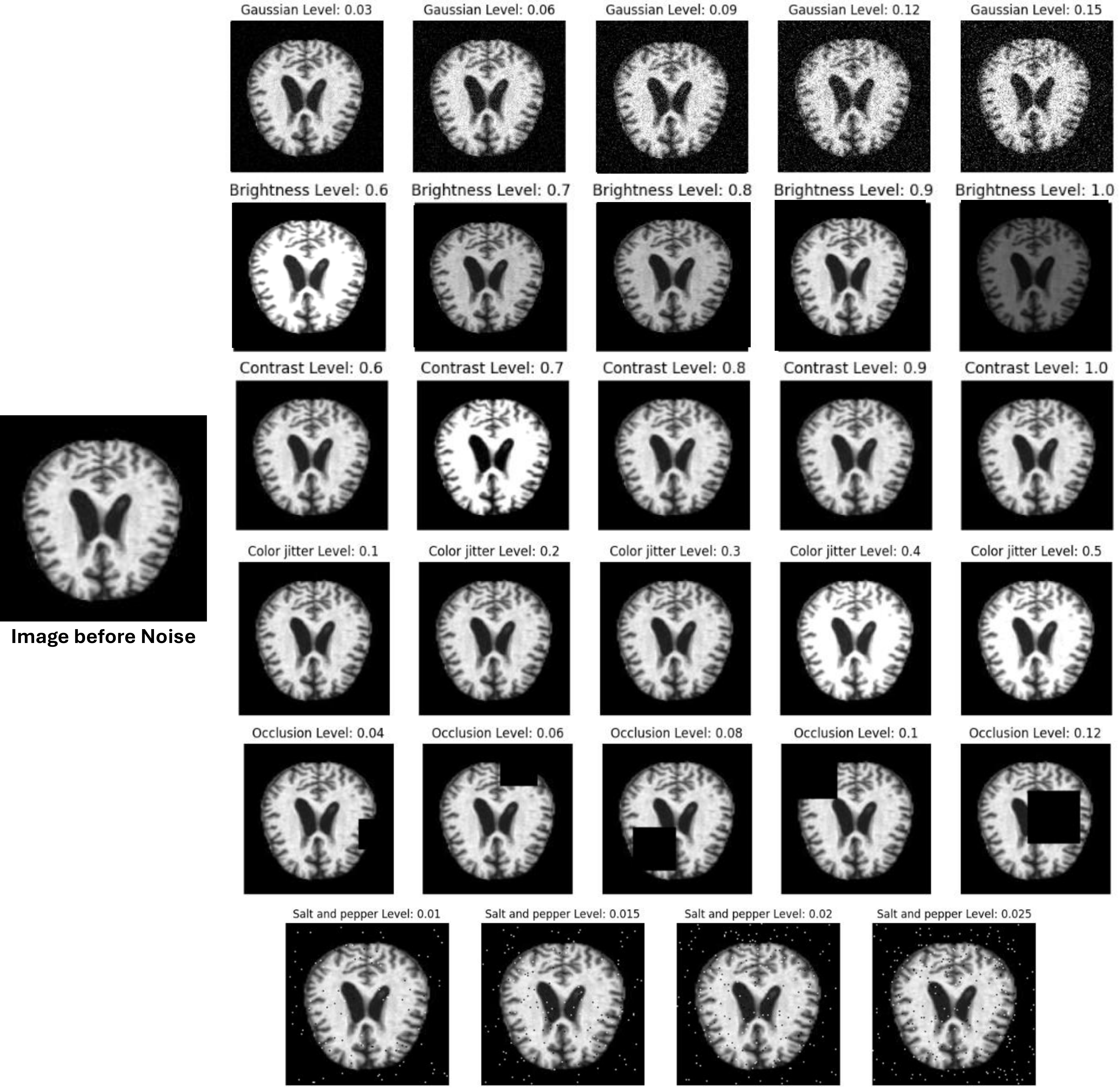}
    \caption{Visualization of various noise and perturbation levels applied to the dataset, including Gaussian noise, brightness, contrast, color jitter, occlusion, and salt and pepper noise. Each row demonstrates the effect of increasing the intensity of the perturbation on the input image.}
        \label{fig:perturbation_variations}
\end{figure*}

\begin{figure*}[htbp]
    \centering
    \includegraphics[width=\textwidth]{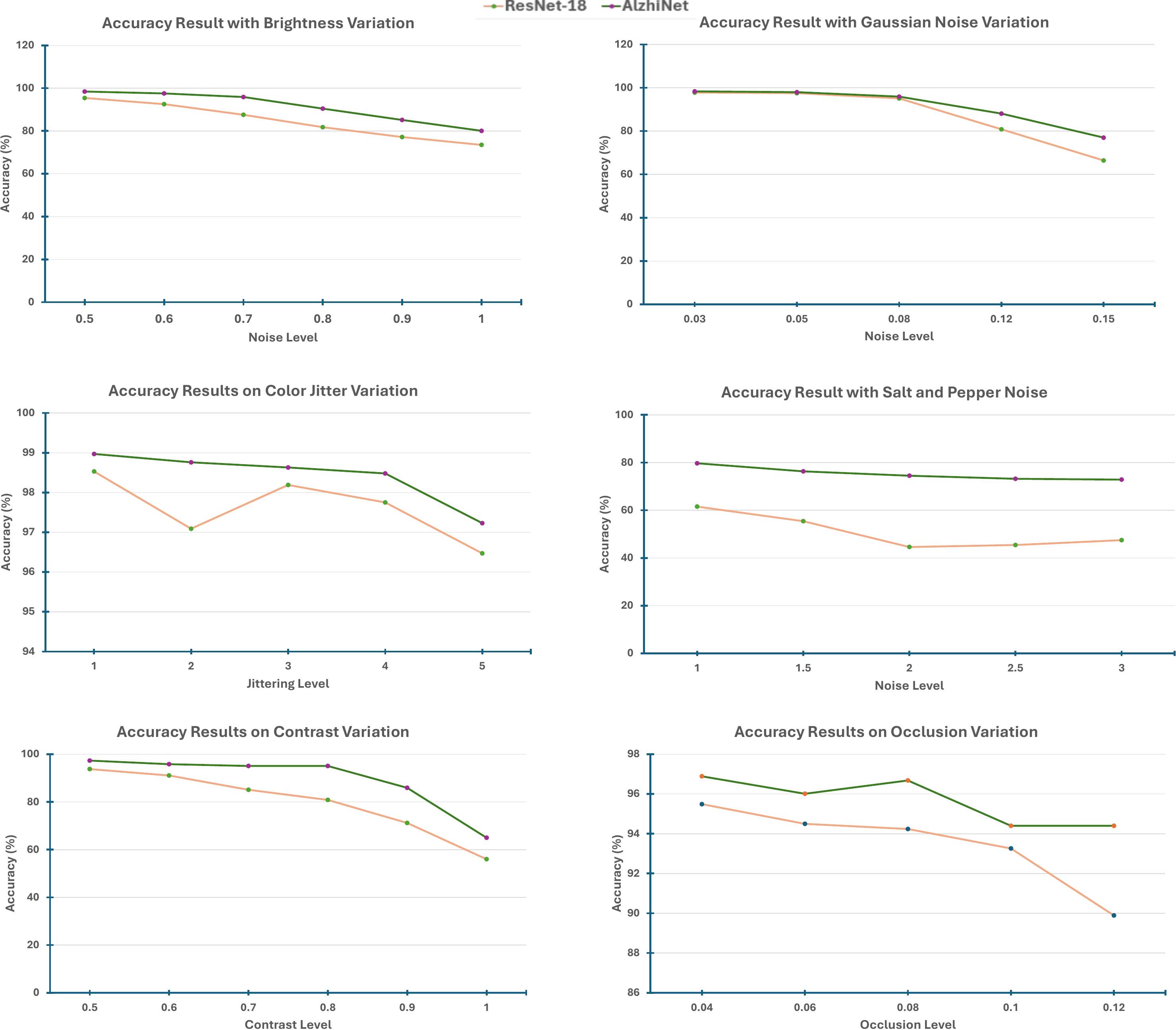} 
    \caption{Accuracy Results with Different Noise and Perturbation Variations on ResNet-18 and AlzhiNet Models}
    \label{fig:accuracy_variations}
\end{figure*}

To test further test robustness of our proposed approach, we conducted an ablation study on AlzhiNet under various perturbation scenarios, including Gaussian noise, brightness, contrast, salt and pepper noise, colour jitter, and occlusion. A visualisation of image sample under different pertubations is presented in Fig \ref{fig:perturbation_variations}. These ablation experiments are essential to understanding the resilience of the model to real-world distortions that may occur in medical imaging data, such as the AlzhiNet. A visualization of the accuracy plot is presented in Fig. \ref{fig:accuracy_variations}. For this experiment, we tested only on the Alzheimer's Kaggle dataset.

For each type of perturbation, we applied a range of intensities to the dataset input before passing it on to either the AlzhiNet or ResNet-18 model and evaluated their performance across key metrics: accuracy, precision, recall, F1 score, specificity, and the AUC score. The results of these experiments are presented in Table \ref{tab:ablation_results}.

\begin{table*}[htbp]
\centering
\caption{Results of Ablation Studies on the AlzhiNet and ResNet-18 Models}
\scalebox{1.1}{
\begin{tabular}{lcccccc}

\toprule
\textbf{Experiment} & \textbf{Model} & \textbf{Accuracy} & \textbf{Precision} & \textbf{Recall} & \textbf{F1 Score} & \textbf{Specificity} \\
\midrule
Gaussian Noise @ 0.03 & ResNet-18 & 97.98 & 97 & 97 & 98 & 99 \\
 & \textbf{AlzhiNet} & \textbf{98.34} & \textbf{99} & \textbf{99} & \textbf{99} & \textbf{99} \\
\hline
Gaussian Noise @ 0.06 & ResNet-18 & 97.10 & 98 & 95 & 96 & 99 \\
 & \textbf{AlzhiNet} & \textbf{97.87} & \textbf{99} & \textbf{97} & \textbf{98} & \textbf{99} \\
\hline
Gaussian Noise @ 0.09 & ResNet-18 & 95.00 & 97 & 84 & 85 & 88 \\
 & \textbf{AlzhiNet} & \textbf{95.18} & \textbf{97} & \textbf{89} & \textbf{92} & \textbf{98} \\
\hline
Gaussian Noise @ 0.12 & ResNet-18 & 80.85 & 88 & 72 & 76 & 78 \\
 & \textbf{AlzhiNet} & \textbf{87.24} & \textbf{90} & \textbf{80} & \textbf{83} & \textbf{95} \\
\hline
Gaussian Noise @ 0.15 & ResNet-18 & 65.6 & 62 & 52 & 52 & 61 \\
 & \textbf{AlzhiNet} & \textbf{66.44} & \textbf{77} & \textbf{61} & \textbf{56} & \textbf{88} \\
\thickhline
Brightness @ 0.5 & ResNet-18 & 95.44 & 94 & 97 & 95 & 98 \\
 & \textbf{AlzhiNet} & \textbf{98.44} & \textbf{99} & \textbf{99} & \textbf{99} & \textbf{99} \\
\hline
Brightness @ 0.6 & ResNet-18 & 92.53 & 86 & 94 & 89 & 97 \\
 & \textbf{AlzhiNet} & \textbf{97.56} & \textbf{94} & \textbf{98} & \textbf{96} & \textbf{99} \\
\hline
Brightness @ 0.7 & ResNet-18 & 87.60 & 77 & 91 & 79 & 95 \\
 & \textbf{AlzhiNet} & \textbf{95.9} & \textbf{88} & \textbf{96} & \textbf{91} & \textbf{98} \\
\hline
Brightness @ 0.8 & ResNet-18 & 81.79 & 72 & 84 & 71 & 94 \\
 & \textbf{AlzhiNet} & \textbf{90.46} & \textbf{77} & \textbf{91} & \textbf{78} & \textbf{97} \\
\hline
Brightness @ 0.9 & ResNet-18 & 77.19 & 70 & 81 & 67 & 93 \\
 & \textbf{AlzhiNet} & \textbf{85.17} & \textbf{75} & \textbf{87} & \textbf{72} & \textbf{95} \\
\hline
Brightness @ 1.0 & ResNet-18 & 73.50 & 70 & 76 & 64 & 92 \\
 & \textbf{AlzhiNet} & \textbf{80.08} & \textbf{73} & \textbf{80} & \textbf{67} & \textbf{94} \\
\thickhline
Contrast @ 0.5 & ResNet-18 & 97.98 & 97 & 97 & 98 & 99 \\
 & \textbf{AlzhiNet} & \textbf{98.34} & \textbf{99} & \textbf{99} & \textbf{99} & \textbf{99} \\
\hline
Contrast @ 0.6 & ResNet-18 & 97.10 & 98 & 95 & 96 & 99 \\
 & \textbf{AlzhiNet} & \textbf{97.87} & \textbf{99} & \textbf{97} & \textbf{98} & \textbf{99} \\
\hline
Contrast @ 0.7 & ResNet-18 & 95.00 & 97 & 84 & 85 & 88 \\
 & \textbf{AlzhiNet} & \textbf{95.18} & \textbf{97} & \textbf{89} & \textbf{92} & \textbf{98} \\
\hline
Contrast @ 0.8 & ResNet-18 & 80.85 & 88 & 72 & 76 & 78 \\
 & \textbf{AlzhiNet} & \textbf{87.24} & \textbf{90} & \textbf{80} & \textbf{83} & \textbf{95} \\
\hline
Contrast @ 0.9 & ResNet-18 & 65.6 & 62 & 52 & 52 & 61 \\
 & \textbf{AlzhiNet} & \textbf{66.44} & \textbf{77} & \textbf{61} & \textbf{56} & \textbf{88} \\
\hline
Contrast @ 1.0 & ResNet-18 & 50.00 & 50 & 50 & 50 & 50 \\
 & \textbf{AlzhiNet} & \textbf{50.00} & \textbf{50} & \textbf{50} & \textbf{50} & \textbf{50} \\
\thickhline
Salt \& Pepper @ 1\% & ResNet-18 & 45.90 & 29 & 21 & 18 & 81 \\
 & \textbf{AlzhiNet} & \textbf{50.00} & \textbf{37} & \textbf{25} & \textbf{17} & \textbf{75} \\
\hline
Salt \& Pepper @ 1.5\% & ResNet-18 & 44.50 & 29 & 20 & 16 & 80 \\
 & \textbf{AlzhiNet} & \textbf{49.84} & \textbf{37} & \textbf{25} & \textbf{17} & \textbf{75} \\
\hline
Salt \& Pepper @ 2.0\% & ResNet-18 & 45.59 & 29 & 21 & 17 & 79 \\
 & \textbf{AlzhiNet} & \textbf{49.84} & \textbf{37} & \textbf{25} & \textbf{17} & \textbf{75} \\
\hline
Salt \& Pepper @ 2.5\% & ResNet-18 & 47.10 & 29 & 23 & 17 & 78 \\
 & \textbf{AlzhiNet} & \textbf{49.95} & \textbf{37} & \textbf{25} & \textbf{17} & \textbf{75} \\
\thickhline
Color Jitter @ 10\% & ResNet-18 & 98.29 & 99 & 98 & 98 & 99 \\
 & \textbf{AlzhiNet} & \textbf{98.81} & \textbf{99} & \textbf{99} & \textbf{99} & \textbf{99} \\
\hline
Color Jitter @ 20\% & ResNet-18 & 97.72 & 99 & 97 & 98 & 99 \\
 & \textbf{AlzhiNet} & \textbf{98.76} & \textbf{99} & \textbf{99} & \textbf{99} & \textbf{99} \\
\hline
Color Jitter @ 30\% & ResNet-18 & 96.78 & 98 & 97 & 99 & 99 \\
 & \textbf{AlzhiNet} & \textbf{98.55} & \textbf{99} & \textbf{99} & \textbf{99} & \textbf{99} \\
\hline
Color Jitter @ 40\% & ResNet-18 & 94.71 & 95 & 94 & 98 & 98 \\
 & \textbf{AlzhiNet} & \textbf{98.13} & \textbf{98} & \textbf{98} & \textbf{98} & \textbf{99} \\
\hline
Color Jitter @ 50\% & ResNet-18 & 90.92 & 82 & 93 & 85 & 97 \\
 & \textbf{AlzhiNet} & \textbf{97.3} & \textbf{92} & \textbf{97} & \textbf{94} & \textbf{99} \\
\thickhline
Occlusion @ 0.04 & ResNet-18 & 95.49 & 95 & 96 & 95 & 98 \\
 & \textbf{AlzhiNet} & \textbf{96.89} & \textbf{98} & \textbf{97} & \textbf{99} & \textbf{100} \\
\hline
Occlusion @ 0.06 & ResNet-18 & 94.50 & 93 & 95 & 94 & 98 \\
 & \textbf{AlzhiNet} & \textbf{96.01} & \textbf{94} & \textbf{96} & \textbf{95} & \textbf{98} \\
 \hline
 Occlusion @ 0.08 & ResNet-18 & 94.24 & 89 & 96 & 92 & 98 \\
 & \textbf{AlzhiNet} & \textbf{96.68} & \textbf{97} & \textbf{97} & \textbf{97} & \textbf{99} \\
 \hline
 Occlusion @ 0.1 & ResNet-18 & 93.26 & 87 & 95 & 90 & 97 \\
 & \textbf{AlzhiNet} & \textbf{94.4} & \textbf{90} & \textbf{95} & \textbf{92} & \textbf{98} \\
 \hline
  Occlusion @ 0.12 & ResNet-18 & 89.89 & 80 & 93 & 85 & 96 \\
 & \textbf{AlzhiNet} & \textbf{94.4} & \textbf{90} & \textbf{95} & \textbf{92} & \textbf{98} \\
 \bottomrule
\end{tabular}
}
\label{tab:ablation_with_noise}
\end{table*}

\subsubsection{Gaussian Noise}
Gaussian noise was introduced at varying levels (0.03 to 0.15) to assess how well the models could handle random noise in the data. As shown in the table, AlzhiNet consistently outperformed ResNet-18 across all noise levels. Notably, AlzhiNet achieved the highest accuracy of 98.34\% at a noise level of 0.03, with precision, recall, F1 score, and specificity all at 99\%. As the noise intensity increased, both models experienced a decrease in performance, but AlzhiNet maintained a higher level of robustness, retaining an accuracy of 66.44\% even at the highest noise level, compared to ResNet-18's 65.6\%.

\subsubsection{Brightness Adjustment}
We adjusted the brightness of the input images in increments from 0.5 to 1.0 to test the models' ability to handle varying lighting conditions. AlzhiNet again demonstrated superior performance, particularly at lower brightness levels. For instance, at a brightness level of 0.5, AlzhiNet achieved an accuracy of 98.44\% with precision, recall, and F1 score all at 99\%. ResNet-18, on the other hand, showed a noticeable drop in performance as brightness increased, particularly at the highest brightness level, where its accuracy fell to 73.50\%.

\subsubsection{Contrast Adjustment}
To evaluate the models under different contrast conditions, the contrast was adjusted from 0.5 to 1.0. AlzhiNet showed strong performance across the board, with the best results observed at lower contrast levels. For example, at a contrast level of 0.5, AlzhiNet achieved an accuracy of 98.34\% with all other metrics close to or at 99\%. In comparison, ResNet-18's performance was slightly lower, especially at the highest contrast level, where it recorded an accuracy of 65.6\%.

\subsubsection{Salt and Pepper Noise}
Salt and pepper noise, which simulates data corruption by randomly flipping pixels, was applied at levels ranging from 1\% to 2.5\%. Both models struggled significantly with this type of noise, but AlzhiNet maintained better overall performance. For example, at a noise level of 1\%, AlzhiNet recorded an accuracy of 50.00\% with a specificity of 75\%, compared to ResNet-18's 45.9\% accuracy and 81\% specificity. This indicates that while both models are sensitive to salt and pepper noise, AlzhiNet is slightly more resilient.

\subsubsection{Colour Jitter}
Colour jitter was introduced at varying intensities (10\% to 50\%) to test the model's robustness to changes in hue, saturation, and brightness. AlzhiNet performed exceptionally well under these conditions, particularly at lower levels of jitter. At a 10\% jitter level, AlzhiNet achieved an accuracy of 98.81\% with perfect or near-perfect scores across all other metrics. ResNet-18 also performed well but showed a larger decline in performance as the jitter intensity increased, particularly at the 50\% level.

\subsubsection{Occlusion}
Finally, The occlusion tests involved masking out portions of the input images with varying levels of occlusion (ranging from 4\% to 12\%). As seen from the results, both models showed some degradation in performance with increasing occlusion, but AlzhiNet demonstrated greater resilience compared to ResNet-18. At a 4\% occlusion level, AlzhiNet achieved an accuracy of 96.89\%, outperforming ResNet-18, which scored 95.49\%. As the occlusion increased, AlzhiNet maintained a higher performance, with an accuracy of 94.4\% even at 12\% occlusion. In contrast, ResNet-18 exhibited a more substantial drop in performance, reaching 89.89\% accuracy at the same occlusion level.

\subsubsection{Summary of Findings}
Overall, the experiments on testing with noises clearly indicate that AlzhiNet is more robust to various perturbations compared to ResNet-18. AlzhiNet consistently delivered better performance across all types of noise and distortions, making it a more reliable choice for real-world applications where medical imaging data may be subject to various forms of degradation. The bolded results in Table \ref{tab:ablation_results} highlight AlzhiNet's superior performance across multiple metrics and perturbation types.

\section{Conclusion}

In this study, we introduced AlzhiNet, a hybrid deep learning model for the classification of Alzheimer's disease from medical imaging data (Brain sMRI). By integrating 2D and 3D CNN architectures, AlzhiNet effectively leverages both spatial and volumetric features, resulting in improved classification accuracy. Our approach is further enhanced by a custom loss function that ensures consistency between the outputs of the 2D and 3D modules.

The results of our extensive experiments demonstrate the efficacy of the AlzhiNet model. The hybrid architecture outperforms individual 2D and 3D models, highlighting the importance of combining these complementary data representations. The ablation studies further reveal the critical role of the custom loss function, particularly the inclusion of the mean squared error term, in maintaining consistency between the model's predictions and enhancing overall performance.

Moreover, the results indicate that the depth and quality of the 3D volumes, generated from augmented 2D slices, significantly impact model accuracy. Careful selection of the weighting factors in the hybrid prediction also plays a vital role in achieving optimal performance.

In summary, AlzhiNet demonstrates robust performance across different datasets, with the hybrid model achieving state-of-the-art results on both the Kaggle and MIRIAD datasets. Our findings underscore the value of integrating 2D and 3D CNNs in medical imaging tasks and suggest that future work could explore additional modalities or even more sophisticated hybrid architectures. The success of AlzhiNet points toward promising avenues for the development of more effective tools for the early diagnosis and treatment of Alzheimer's disease, ultimately contributing to better patient outcomes.

Future research could extend this work by exploring the integration of additional imaging modalities or employing more advanced network architectures to further improve classification accuracy. Additionally, the application of AlzhiNet to other neurodegenerative diseases could provide valuable insights and contribute to the broader field of medical image analysis.

\section*{Acknowledgment}

 This work was supported by the National Natural Science Foundation of China (No.42075129), Hebei Province Natural Science Foundation (No.E2021202179), Key Research and Development Project from Hebei Province (No.21351803D), and Hebei special project for key technology and product R\&D (No.SJMYF2022Y06).


%





\ifCLASSOPTIONcaptionsoff
  \newpage
\fi





\bibliographystyle{IEEEtran}
\bibliography{IEEEabrv, main}

\begin{thebibliography}{10}
\providecommand{\url}[1]{#1}
\csname url@rmstyle\endcsname
\providecommand{\newblock}{\relax}
\providecommand{\bibinfo}[2]{#2}
\providecommand\BIBentrySTDinterwordspacing{\spaceskip=0pt\relax}
\providecommand\BIBentryALTinterwordstretchfactor{4}
\providecommand\BIBentryALTinterwordspacing{\spaceskip=\fontdimen2\font plus
\BIBentryALTinterwordstretchfactor\fontdimen3\font minus \fontdimen4\font\relax}
\providecommand\BIBforeignlanguage[2]{{%
\expandafter\ifx\csname l@#1\endcsname\relax
\typeout{** WARNING: IEEEtran.bst: No hyphenation pattern has been}%
\typeout{** loaded for the language `#1'. Using the pattern for}%
\typeout{** the default language instead.}%
\else
\language=\csname l@#1\endcsname
\fi
#2}}

\bibitem{rasmussen2019alzheimer}
J.~Rasmussen and H.~Langerman, ``Alzheimer’s disease--why we need early diagnosis,'' \emph{Degenerative neurological and neuromuscular disease}, pp. 123--130, 2019.

\bibitem{fulton2019classification}
L.~V. Fulton, D.~Dolezel, J.~Harrop, Y.~Yan, and C.~P. Fulton, ``Classification of alzheimer’s disease with and without imagery using gradient boosted machines and resnet-50,'' \emph{Brain sciences}, vol.~9, no.~9, p. 212, 2019.

\bibitem{lian2018hierarchical}
C.~Lian, M.~Liu, J.~Zhang, and D.~Shen, ``Hierarchical fully convolutional network for joint atrophy localization and alzheimer's disease diagnosis using structural mri,'' \emph{IEEE transactions on pattern analysis and machine intelligence}, vol.~42, no.~4, pp. 880--893, 2018.

\bibitem{zhang2021explainable}
X.~Zhang, L.~Han, W.~Zhu, L.~Sun, and D.~Zhang, ``An explainable 3d residual self-attention deep neural network for joint atrophy localization and alzheimer’s disease diagnosis using structural mri,'' \emph{IEEE journal of biomedical and health informatics}, vol.~26, no.~11, pp. 5289--5297, 2021.

\bibitem{aberathne2023detection}
I.~Aberathne, D.~Kulasiri, and S.~Samarasinghe, ``Detection of alzheimer’s disease onset using mri and pet neuroimaging: longitudinal data analysis and machine learning,'' \emph{Neural Regeneration Research}, vol.~18, no.~10, pp. 2134--2140, 2023.

\bibitem{wen2020convolutional}
J.~Wen, E.~Thibeau-Sutre, M.~Diaz-Melo, J.~Samper-Gonz{\'a}lez, A.~Routier, S.~Bottani, D.~Dormont, S.~Durrleman, N.~Burgos, O.~Colliot, \emph{et~al.}, ``Convolutional neural networks for classification of alzheimer's disease: Overview and reproducible evaluation,'' \emph{Medical image analysis}, vol.~63, p. 101694, 2020.

\bibitem{chan2020deep}
H.-P. Chan, R.~K. Samala, L.~M. Hadjiiski, and C.~Zhou, ``Deep learning in medical image analysis,'' \emph{Deep learning in medical image analysis: challenges and applications}, pp. 3--21, 2020.

\bibitem{wang2023hybrid}
X.~Wang and Z.~Liang, ``Hybrid network model based on 3d convolutional neural network and scalable graph convolutional network for hyperspectral image classification,'' \emph{IET Image Processing}, vol.~17, no.~1, pp. 256--273, 2023.

\bibitem{yeoh2023transfer}
P.~S.~Q. Yeoh, K.~W. Lai, S.~L. Goh, K.~Hasikin, X.~Wu, and P.~Li, ``Transfer learning-assisted 3d deep learning models for knee osteoarthritis detection: Data from the osteoarthritis initiative,'' \emph{Frontiers in Bioengineering and Biotechnology}, vol.~11, p. 1164655, 2023.

\bibitem{klaiber2021systematic}
M.~Klaiber, D.~Sauter, H.~Baumgartl, and R.~Buettner, ``A systematic literature review on transfer learning for 3d-cnns,'' in \emph{2021 international joint conference on neural networks (IJCNN)}.\hskip 1em plus 0.5em minus 0.4em\relax IEEE, 2021, pp. 1--10.

\bibitem{shakarami2020cad}
A.~Shakarami, H.~Tarrah, and A.~Mahdavi-Hormat, ``A cad system for diagnosing alzheimer’s disease using 2d slices and an improved alexnet-svm method,'' \emph{Optik}, vol. 212, p. 164237, 2020.

\bibitem{qiu2020development}
S.~Qiu, P.~S. Joshi, M.~I. Miller, C.~Xue, X.~Zhou, C.~Karjadi, G.~H. Chang, A.~S. Joshi, B.~Dwyer, S.~Zhu, \emph{et~al.}, ``Development and validation of an interpretable deep learning framework for alzheimer’s disease classification,'' \emph{Brain}, vol. 143, no.~6, pp. 1920--1933, 2020.

\bibitem{amin2019alzheimer}
M.~Amin-Naji, H.~Mahdavinataj, and A.~Aghagolzadeh, ``Alzheimer's disease diagnosis from structural mri using siamese convolutional neural network,'' in \emph{2019 4th international conference on pattern recognition and image analysis (IPRIA)}.\hskip 1em plus 0.5em minus 0.4em\relax IEEE, 2019, pp. 75--79.

\bibitem{jabason2019classification}
E.~Jabason, M.~O. Ahmad, and M.~Swamy, ``Classification of alzheimer’s disease from mri data using an ensemble of hybrid deep convolutional neural networks,'' in \emph{2019 IEEE 62nd international Midwest symposium on circuits and systems (MWSCAS)}.\hskip 1em plus 0.5em minus 0.4em\relax IEEE, 2019, pp. 481--484.

\bibitem{odusami2021comparable}
M.~Odusami, R.~Maskeliunas, R.~Dama{\v{s}}evi{\v{c}}ius, and S.~Misra, ``Comparable study of pre-trained model on alzheimer disease classification,'' in \emph{Computational Science and Its Applications--ICCSA 2021: 21st International Conference, Cagliari, Italy, September 13--16, 2021, Proceedings, Part V 21}.\hskip 1em plus 0.5em minus 0.4em\relax Springer, 2021, pp. 63--74.

\bibitem{he2016deep}
K.~He, X.~Zhang, S.~Ren, and J.~Sun, ``Deep residual learning for image recognition,'' in \emph{Proceedings of the IEEE conference on computer vision and pattern recognition}, 2016, pp. 770--778.

\bibitem{yu2019abnormality}
X.~Yu and S.-H. Wang, ``Abnormality diagnosis in mammograms by transfer learning based on resnet18,'' \emph{Fundamenta Informaticae}, vol. 168, no. 2-4, pp. 219--230, 2019.

\bibitem{odusami2021analysis}
M.~Odusami, R.~Maskeli{\=u}nas, R.~Dama{\v{s}}evi{\v{c}}ius, and T.~Krilavi{\v{c}}ius, ``Analysis of features of alzheimer’s disease: Detection of early stage from functional brain changes in magnetic resonance images using a finetuned resnet18 network,'' \emph{Diagnostics}, vol.~11, no.~6, p. 1071, 2021.

\bibitem{odusami2021resd}
M.~Odusami, R.~Maskeli{\=u}nas, R.~Dama{\v{s}}evi{\v{c}}ius, and S.~Misra, ``Resd hybrid model based on resnet18 and densenet121 for early alzheimer disease classification,'' in \emph{International Conference on Intelligent Systems Design and Applications}.\hskip 1em plus 0.5em minus 0.4em\relax Springer, 2021, pp. 296--305.

\bibitem{liang2021alzheimer}
G.~Liang, X.~Xing, L.~Liu, Y.~Zhang, Q.~Ying, A.-L. Lin, and N.~Jacobs, ``Alzheimer’s disease classification using 2d convolutional neural networks,'' in \emph{2021 43rd annual international conference of the ieee engineering in medicine \& biology society (embc)}.\hskip 1em plus 0.5em minus 0.4em\relax IEEE, 2021, pp. 3008--3012.

\bibitem{nawaz2020deep}
A.~Nawaz, S.~M. Anwar, R.~Liaqat, J.~Iqbal, U.~Bagci, and M.~Majid, ``Deep convolutional neural network based classification of alzheimer's disease using mri data,'' in \emph{2020 IEEE 23rd International Multitopic Conference (INMIC)}.\hskip 1em plus 0.5em minus 0.4em\relax IEEE, 2020, pp. 1--6.

\bibitem{ebrahimi2021deep}
A.~Ebrahimi, S.~Luo, R.~Chiong, A.~D.~N. Initiative, \emph{et~al.}, ``Deep sequence modelling for alzheimer's disease detection using mri,'' \emph{Computers in Biology and Medicine}, vol. 134, p. 104537, 2021.

\bibitem{parmar2020deep}
H.~S. Parmar, B.~Nutter, R.~Long, S.~Antani, and S.~Mitra, ``Deep learning of volumetric 3d cnn for fmri in alzheimer’s disease classification,'' in \emph{Medical imaging 2020: Biomedical applications in molecular, structural, and functional imaging}, vol. 11317.\hskip 1em plus 0.5em minus 0.4em\relax SPIE, 2020, pp. 66--71.

\bibitem{yang2018visual}
C.~Yang, A.~Rangarajan, and S.~Ranka, ``Visual explanations from deep 3d convolutional neural networks for alzheimer’s disease classification,'' in \emph{AMIA annual symposium proceedings}, vol. 2018.\hskip 1em plus 0.5em minus 0.4em\relax American Medical Informatics Association, 2018, p. 1571.

\bibitem{qiao2021fusion}
H.~Qiao, L.~Chen, and F.~Zhu, ``A fusion of multi-view 2d and 3d convolution neural network based mri for alzheimer’s disease diagnosis,'' in \emph{2021 43rd Annual International Conference of the IEEE Engineering in Medicine \& Biology Society (EMBC)}.\hskip 1em plus 0.5em minus 0.4em\relax IEEE, 2021, pp. 3317--3321.

\bibitem{hosseini2016alzheimer}
E.~Hosseini-Asl, R.~Keynton, and A.~El-Baz, ``Alzheimer's disease diagnostics by adaptation of 3d convolutional network,'' in \emph{2016 IEEE International Conference on image processing (ICIP)}.\hskip 1em plus 0.5em minus 0.4em\relax IEEE, 2016, pp. 126--130.

\bibitem{dubey2019alzheimer}
S.~Dubey, ``Alzheimer’s dataset (4 class of images),'' \emph{Kaggle, Dec}, vol.~26, 2019.

\bibitem{malone2013miriad}
I.~B. Malone, D.~Cash, G.~R. Ridgway, D.~G. MacManus, S.~Ourselin, N.~C. Fox, and J.~M. Schott, ``Miriad—public release of a multiple time point alzheimer's mr imaging dataset,'' \emph{NeuroImage}, vol.~70, pp. 33--36, 2013.

\bibitem{med2image}
FNNDSC, ``med2image,'' \url{https://github.com/FNNDSC/med2image}, 2020, accessed: 2023-10-01.

\bibitem{prasanna2024ad}
G.~Prasanna, ``Ad-tl: Alzheimer’s disease prediction using transfer learning,'' \emph{J. Electrical Systems}, vol.~20, no.~6s, pp. 1132--1147, 2024.

\bibitem{ahmed2022dad}
G.~Ahmed, M.~J. Er, M.~M.~S. Fareed, S.~Zikria, S.~Mahmood, J.~He, M.~Asad, S.~F. Jilani, and M.~Aslam, ``Dad-net: Classification of alzheimer’s disease using adasyn oversampling technique and optimized neural network,'' \emph{Molecules}, vol.~27, no.~20, p. 7085, 2022.

\bibitem{abunadi2022deep}
I.~Abunadi, ``Deep and hybrid learning of mri diagnosis for early detection of the progression stages in alzheimer’s disease,'' \emph{Connection Science}, vol.~34, no.~1, pp. 2395--2430, 2022.

\bibitem{sharma2022deep}
S.~Sharma, K.~Guleria, S.~Tiwari, and S.~Kumar, ``A deep learning based convolutional neural network model with vgg16 feature extractor for the detection of alzheimer disease using mri scans,'' \emph{Measurement: Sensors}, vol.~24, p. 100506, 2022.

\bibitem{loddo2022deep}
A.~Loddo, S.~Buttau, and C.~Di~Ruberto, ``Deep learning based pipelines for alzheimer's disease diagnosis: a comparative study and a novel deep-ensemble method,'' \emph{Computers in biology and medicine}, vol. 141, p. 105032, 2022.

\bibitem{el2023accurate}
A.~A.~A. El-Latif, S.~A. Chelloug, M.~Alabdulhafith, and M.~Hammad, ``Accurate detection of alzheimer’s disease using lightweight deep learning model on mri data,'' \emph{Diagnostics}, vol.~13, no.~7, p. 1216, 2023.

\bibitem{chabib2023deepcurvmri}
C.~M. Chabib, L.~J. Hadjileontiadis, and A.~Al~Shehhi, ``Deepcurvmri: Deep convolutional curvelet transform-based mri approach for early detection of alzheimer’s disease,'' \emph{IEEE Access}, vol.~11, pp. 44\,650--44\,659, 2023.

\bibitem{de2023prediction}
K.~De~Silva and H.~Kunz, ``Prediction of alzheimer's disease from magnetic resonance imaging using a convolutional neural network,'' \emph{Intelligence-Based Medicine}, vol.~7, p. 100091, 2023.

\bibitem{hassan2024residual}
N.~Hassan, A.~S. Musa~Miah, and J.~Shin, ``Residual-based multi-stage deep learning framework for computer-aided alzheimer’s disease detection,'' \emph{Journal of Imaging}, vol.~10, no.~6, p. 141, 2024.

\bibitem{alsaeed2022brain}
D.~AlSaeed and S.~F. Omar, ``Brain mri analysis for alzheimer’s disease diagnosis using cnn-based feature extraction and machine learning,'' \emph{Sensors}, vol.~22, no.~8, p. 2911, 2022.

\bibitem{silva2019model}
I.~R. Silva, G.~S. Silva, R.~G. de~Souza, W.~P. dos Santos, and R.~A. d.~A. Fagundes, ``Model based on deep feature extraction for diagnosis of alzheimer’s disease,'' in \emph{2019 international joint conference on neural networks (IJCNN)}.\hskip 1em plus 0.5em minus 0.4em\relax IEEE, 2019, pp. 1--7.

\end{thebibliography}
%


\begin{IEEEbiography}[{\includegraphics[width=1in,height=1.8in,clip,keepaspectratio]{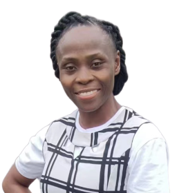}}]{Romoke Grace Akindele}
received the B.Tech. degree in pure and applied physics from the Ladoke Akintola University of Technology (LAUTECH), Ogbomoso, Nigeria, in 2010, and the M.Tech. Degree in communication physics from the Federal University of Technology Akure (FUTA), Nigeria, in 2016. She is currently pursuing a Ph.D. degree in electronics and information engineering at the Hebei University of Technology, China. Her research interests include wireless communication, image processing,  deep learning, and computer vision.
\end{IEEEbiography}

\begin{IEEEbiography}[{\includegraphics[width=1in,height=1.8in,clip,keepaspectratio]{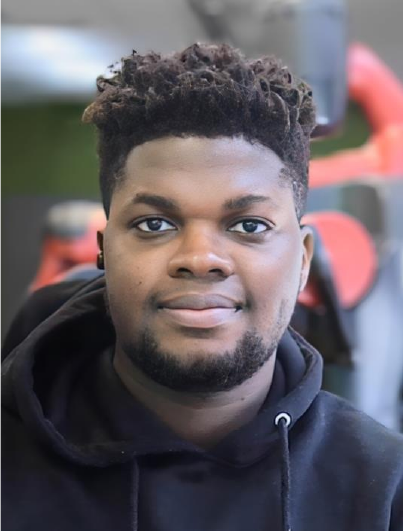}}]{Samuel Adebayo}
earned a First Class BSc in Electronic and Computer Engineering from Lagos State University, Lagos, Nigeria (2016-2020) and PhD in Machine Learning at Queen's University Belfast, UK (2020-2024). Samuel's research is deeply rooted in computational intelligence, primarily focusing on the intricate dynamics of computational intelligence and its broader application. This aligns with his interest in advancing the fields of Computer Vision, deep Learning, Natural Language Processing, and Bayesian Machine Learning. He is currently a Research Fellow at Queen's University Belfast.
\end{IEEEbiography}

\begin{IEEEbiography}[{\includegraphics[width=1in,height=1.8in,clip,keepaspectratio]{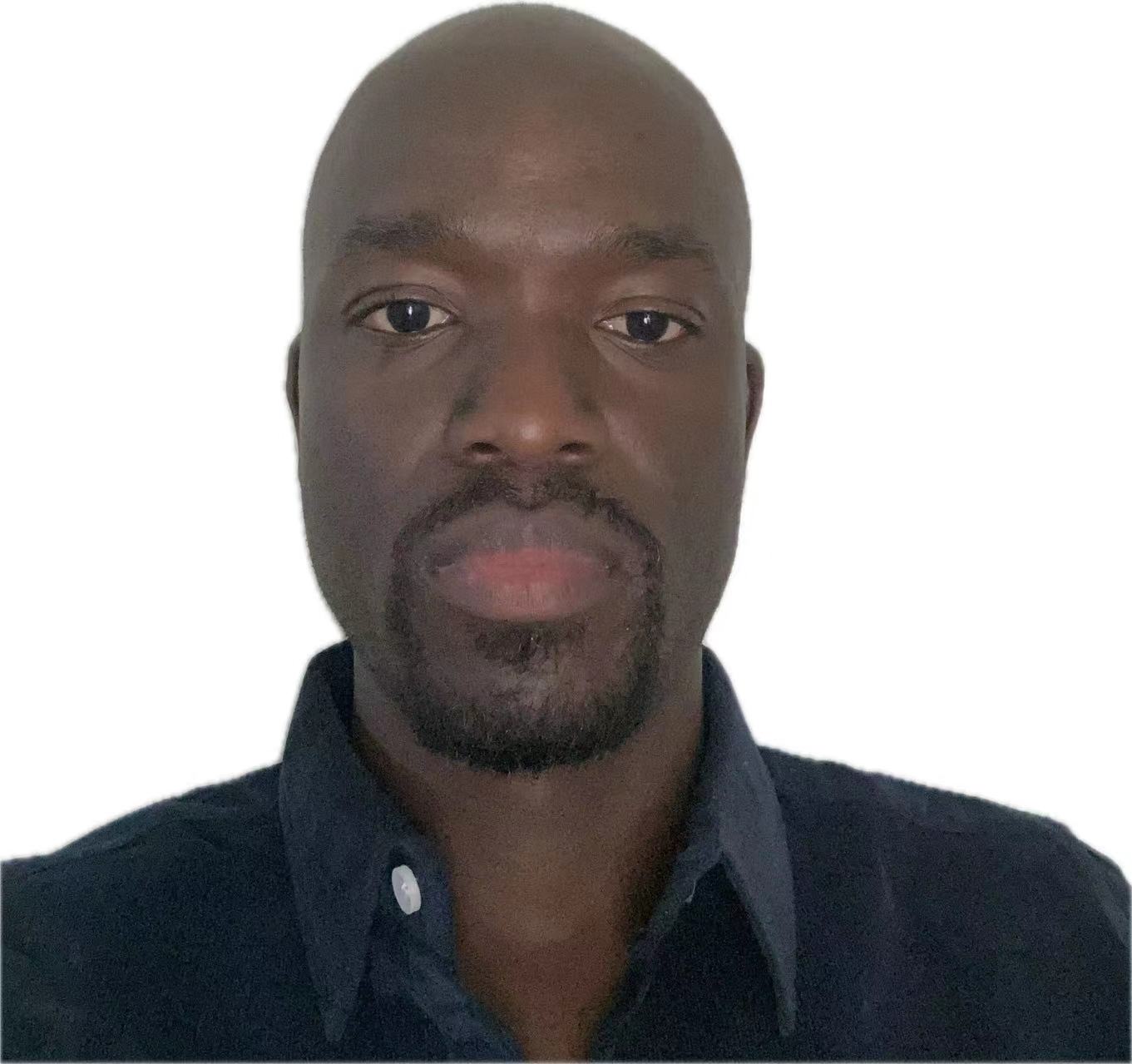}}]{Paul Shekonya Kanda}
received the B.Eng. degree in electrical and computer engineering from the Federal University of Technology Minna (FUTM), Nigeria, in 2011, and the M.Tech. degree in computer science and technology from the Liaoning University of Technology (LUT), China, in 2016. He is currently pursuing a Ph.D. degree in electronics and information engineering with the Hebei University of Technology, Tianjin, China. His research interests include deep learning and image processing.
\end{IEEEbiography}

\begin{IEEEbiography}[{\includegraphics[width=1in,height=1.8in,clip,keepaspectratio]{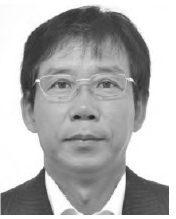}}]{Ming Yu}
(Member, IEEE) received a Ph.D. degree in communication and information systems from the Beijing Institute of Technology. Since 1989, he has been a teacher at the Hebei University of Technology and has also been a full professor at the School of Computer Science and Engineering since 2000. His research interests include biometrics of voice and image vision information fusion, image mathematical transformation, an efficient algorithm of image and video coding, computer networks, and high-compression image video transmission under wireless mobile networks.
\end{IEEEbiography}





\vfill


\end{document}